\documentclass[showpacs,amsmath,amssymb]{revtex4}

\usepackage{graphicx}
\usepackage{dcolumn}
\usepackage{bm}
\usepackage{amsmath,epsfig}

\newcommand{\ket}[1]{\left\vert#1\right\rangle}
\newcommand{\bra}[1]{\left\langle#1\right\vert}
\newcommand{\braket}[2]{\left\langle#1\vert#2\right\rangle}

\begin{document}

\title{Translational dynamics effects on the non-local correlations between two atoms}
\author{M. Tumminello, A. Vaglica, G. Vetri}
\affiliation{
  Istituto Nazionale di Fisica della Materia and Dipartimento di Scienze Fisiche ed Astronomiche, Universit\`{a} di Palermo -
  via Archirafi 36, 90123 Palermo, Italy} 

\date{\today}

\begin{abstract}
A pair of  atoms interacting successively with the field of the
same cavity and exchanging a single photon, leave the cavity in an
entangled state of Einstein-Podolsky-Rosen (EPR) type (see, for
example, [S.J.D. Phoenix, and S.M. Barnett, J. Mod. Opt.
\textbf{40} (1993) 979]). By implementing the model with the
translational degrees of freedom, we show in this letter that the
entanglement with the translational atomic variables can lead,
under appropriate conditions, towards the separability of the
internal variables of the two atoms. This implies that the
translational dynamics can lead, in some cases, to difficulties in
observing the Bell's inequality violation for massive particles.
\end{abstract}

\maketitle

A necessary condition for attaining fruitful teleportation
\cite{tel} is the possibility of constructing ``non locally''
correlated systems. Recently, it has been payed attention to
teleportation of massive particles and, more generally, to
non-local correlations, separability and related issues \cite{Har,
Ita, Phoe, Kni, Fre, Zub, Ray}. A simple model which can realize
an EPR state consists of two atoms which interact successively
with the field of an optical cavity. As a consequence of the
entanglement developed during the interaction between the first
atom and the field, quantum correlations between the two atoms do
arise as the second one interacts with the field of the same
cavity. Using the standard Jaynes-Cummings (JC) model, non-local
correlations have been predicted \cite{Phoe} which can lead to a
violation of the Bell's inequality.\\
In this letter we suggest that a careful analysis of the
interatomic correlations may require the quantization of the
translational dynamics of the two atoms along the cavity axis. In
fact, it is here shown that the quantization of the translational
dynamics affects the non-local features of the interatomic
correlations, at least when the atoms enter the cavity in a region
where the field gradient is different from zero (for example in a
nodal region). To take into account these translational effects,
we adopt the optical Stern-Gerlach (SG) model \cite{sgo} for the
atom-field interaction, while to investigate on the non-local
features we test the Bell's inequality \cite{Asp} and the
separability \cite{Per,Wer} for the reduced density matrix
describing the internal degrees of freedom of the two atoms, after
their interaction with the cavity. For this system our main
quantitative result reveals the appearance of damping terms in the
correlation functions of the two-atom internal variables, induced
by the entanglement with the translational dynamics. These damping
factors destroy the quantum nature of the correlations
just for an interaction time of a few Rabi oscillations.\\
Let us consider a system composed by two two-level atoms
interacting, not simultaneously, with the e.m. field of  the same
undamped cavity. At the time $t=0$, the first atom, say $A_{1}$,
enters the cavity moving prevalently along the z-direction,
orthogonal to the x-cavity axis, and interacts with the field for
a time $t_{1}$. We assume that the atomic velocity along the
z-direction is large enough to treat classically this component of
the motion. At a later time $t_{2}$, the second atom, say $A_{2}$,
starts interacting with the field-state in the cavity modified by
the first atom. At the time $t_{3}$, $A_{2}$ leaves the cavity and
both the atoms evolve freely. We assume that the atoms enter the
cavity in proximity of a nodal region of the cavity resonant mode
\emph{k}, and the width of their wave packets is sufficiently
small with respect to the wavelength of the \emph{k}-mode. In the
interaction picture, the Hamiltonian of the system at all times
reads
\begin{equation}\label{ham}
  \hat{H}^{I}(t)=\hbar\varepsilon k(\hat{x}_{1}+\frac{\hat{p}_{1}}
  {m}t)\mu_{t}(0,t_{1})\hat{u}_{1}+ \hbar\varepsilon k(\hat{x}_{2}+\frac{\hat{p}_{2}}{m}t)
  \mu_{t}(t_{2},t_{3})\hat{u}_{2},
\end{equation}
\\
where the observable $\hat{x}_{i}$ describes the position of
$A_{i}$ with respect to the nodal point and $\hat{p}_{i}$ is the
conjugate momentum. The atom-field interaction operator
$\hat{u}_{i}=\hat{a}^{\dag}\hat{S}_{-,i}+\hat{a}\hat{S}_{+,i}$
mixes the annihilation and creation field-operators $\hat{a}$ and
$\hat{a}^{\dag}$ with the usual spin $1/2$ operators
$\hat{S}_{\pm,i}$. Finally
$\mu_{t}(x,y)=\theta_{t}(x)-\theta_{t}(y)$ is a linear combination
of the usual step-functions with different points of discontinuity
(x and y), helpful to distinguish the different time-ranges
concerning the successive interaction of the two atoms (with same
mass \emph{m} and same atom-field coupling constant $\varepsilon$)
with the cavity-field. \\
To evaluate the state of the entire system at some time $t \geq
t_{3}$, when both the atoms are outside the cavity, we look at the
evolution operator (for the one atom case see Ref.\cite{Cus})
\begin{eqnarray}\label{evop}
\hat{U}^{I}(t\geq t_{3})=\exp\{-i \varepsilon k \hat{u}_{2}
(t_{3}-t_{2})[\hat{x}_{2}+ \frac{\hat{p}_{2}}{2
m}(t_{3}+t_{2})]\}\cdot\quad\quad\quad\quad
\quad\quad\quad\quad\quad\quad\quad\nonumber \\
\cdot\exp[i \hbar \frac{\varepsilon^{2}
k^{2}}{12m}\hat{u}_{2}^{2}(t_{3}-t_{2})^{3}]\exp[i \hbar
\frac{\varepsilon^{2} k^{2}}{12 m}\hat{u}_{1}^{2}t_{1}^{3}]\exp[-i
\varepsilon k\hat{u}_{1}t_{1}(\hat{x}_{1}+\frac{\hat{p}_{1}}{2
m}t_{1})].
\end{eqnarray}
For ease of comparison with the results obtained in the ambit of
the standard JC model \cite{Phoe}, we consider the simple case in
which, initially, both the two atoms are in the ground state
$\ket{g_{i}}$, and the cavity mode contains just one photon,
\begin{equation}\label{psi0}
\ket{\psi(0)}=\ket{g_{1}}\ket{g_{2}}\ket{1}\ket{\varphi_{1}(0)}\ket{\varphi_{2}(0)},
\end{equation}
where $\ket{\varphi_{i}(0)}$ is a translational state of the atom
$A_{i}$. At a time $t \ge t_{3}$, after the successive atom-field
interactions, the entire system state is
\begin{equation}\label{psi1}
\ket{\psi(t_3)}=[\ket{S_{1}^{-}}\ket{\varphi_{2}(0)}\ket{e_{1}}\ket{g_{2}}+
\ket{S_{1}^{+}}\ket{S_{2}^{-}}\ket{g_{1}}\ket{e_{2}}]\ket{0}+
\ket{S_{1}^{+}}\ket{S_{2}^{+}}\ket{g_{1}}\ket{g_{2}}\ket{1},
\end{equation}
where $\ket{e_{i}}$ indicates the excited state of $A_{i}$. The
atomic translational parts modified because of the interaction
with the cavity-photons are
\begin{equation}\label{psi2}
\ket{S_{i}^{\pm}}=\frac{1}{2}[\ket{\phi_{i}^{+}}\pm\ket{\phi_{i}^{-}}],
\end{equation}
where $i\epsilon \{1,2\}$ and
\begin{eqnarray}
\ket{\phi_{1}^{\pm}}=\exp{[\mp i \varepsilon
k(\hat{x}_{1}+\frac{\hat{p}_{1}}{2 m}t_{1})t_{1}]}\cdot\exp{[i
\hbar \frac{\varepsilon^{2} k^{2}}{12
m}t_{1}^{3}]}\ket{\varphi_{1}(0)},\label{Phi1}\\
\ket{\phi_{2}^{\pm}}=\exp{\{\mp i \varepsilon
k[\hat{x}_{2}+\frac{\hat{p}_{2}}{2
m}(t_{3}+t_{2})](t_{3}-t_{2})\}}\cdot\exp{[i \hbar
\frac{\varepsilon^{2} k^{2}}{12
m}(t_{3}-t_{2})^{3}]}\ket{\varphi_{2}(0)}. \label{Phi2}
\end{eqnarray}
As it is evident, state (\ref{psi1}) exhibits crossed correlations
between the atomic translation, the atomic internal dynamics and
the field variables. Since the aim of this paper is to analyze the
entanglement between the internal variables of the two atoms as a
function of the interaction time, we now trace on the atomic
translational and field variables, and we obtain the following
reduced density matrix
\begin{eqnarray}
\rho=Tr_{f,s_{1},s_{2}}(\ket{\psi(t_{3})}\bra{\psi(t_{3})})=
P_{1}\ket{g_{1}}\ket{g_{2}}\bra{g_{1}}\bra{g_{2}}+\quad\quad\quad\quad\quad\quad\quad\quad
\quad\quad\quad\quad\quad\quad\nonumber\\
+P_{2}\{c_{2}^{2}\ket{e_{1}}\ket{g_{2}}\bra{e_{1}}\bra{g_{2}}
+c_{1}^{2}\ket{g_{1}}\ket{e_{2}}\bra{g_{1}}\bra{e_{2}}
+c_{1}c_{2}[q\ket{e_{1}}\ket{g_{2}}\bra{g_{1}}\bra{e_{2}}+h.c.]\}\label{rhorid}
\end{eqnarray}
where $P_{1}+P_{2}=1$ and
\begin{eqnarray}\label{pc}
P_{1}=\frac{1}{4}(1+c_{R}^{(1)})(1+c_{R}^{(2)}),\quad
c_{1}=\sqrt{\frac{(1+c_{R}^{(1)})(1-c_{R}^{(2)})}{4P_{2}}}, \quad
c_{2}=\sqrt{\frac{(1-c_{R}^{(1)})}{2P_{2}}}.
\end{eqnarray}
Apart from the complex parameter
\begin{equation}\label{q}
q=i\frac{c_{-}-c_{+}}{\sqrt{2(1-c_{R}^{(1)})(1+c_{R}^{(1)})(1-c_{R}^{(2)})}}
c_{I}^{(1)},
\end{equation}
equation(\ref{rhorid}) is formally very similar to the
corresponding equation of Ref.\cite{Phoe}. However, our approach
involves the entanglement with the atomic translation whose
effects are encoded in the coefficients

\begin{equation}\label{param}
c_{R}^{(i)}=Re(\braket{\phi_{i}^{+}}{\phi_{i}^{-}}),\quad
c_{I}^{(i)}=Im(\braket{\phi_{i}^{+}}{\phi_{i}^{-}}),\quad
c_{\pm}=\braket{\phi_{2}^{\pm}}{\varphi_{2}(0)}.
\end{equation}
The scalar products which appear in this equation play a crucial
role in determining the separability of $\rho$. They are
characterized, in fact, by a non dissipative damping term whose
origins go back to the distance in the phase space of the
translational components \cite{Aha, Vag, Chia}. For instance, we
found
\begin{equation}\label{scalar}
  \braket{\phi^{+}_{1}}{\phi^{-}_{1}}=\exp(-\frac{d^{2}}{8})\exp(2
i x_{1} \varepsilon k t_{1}),
\end{equation}
where
\begin{equation}\label{distance}
  d^{2}=\frac{[x_{1}^{+}(t_{1})-x_{1}^{-}(t_{1})]^2}{\sigma_{x_{1}}^{2}}+
  \frac{[p_{1}^{+}(t_{1})-p_{1}^{-}(t_{1})]^2}{\sigma_{p_{1}}^{2}}
\end{equation}
may be interpreted as the square of an adimensional distance in
the phase space defined in terms of $x_{1}^{\pm}(t)=x_{1}\mp a
\,t^{2}/2$ and $p_{1}^{\pm}(t)=\mp m\,a\,t$ (with $a=\frac{\hbar k
\epsilon}{m}$) and measured in units of $\sigma_{x}$ and
$\sigma_{p}$, along $x$ and $p$, respectively. Similar behaviors
hold for the other scalar products. In deriving this and similar
expressions we have considered, for both the atoms, Gaussian
initial packets of minimum uncertainty, centered in $x_{i}$ with
zero mean velocity along the x-cavity axis and of width
$\sigma_{x_{i}}$ ($\sigma_{p_{i}}= \frac{\hbar}{2
\sigma_{x_{i}}}$). If the damping factors involved in the scalar
products of eq. (\ref{param}) are disregarded, we recover the
reduced state of Ref.\cite{Phoe}, obtained in the ambit of the JC
model
\begin{eqnarray}\label{rhoridjc}
\rho=\cos^{4}(\varepsilon_{JC}T)\ket{g_{1}}\ket{g_{2}}\bra{g_{1}}
\bra{g_{2}}+\quad\quad\quad\quad\quad\quad\quad\quad\quad\quad\quad\quad\quad\quad\quad
\quad\quad\quad\quad\quad\quad \nonumber
\\ \quad\quad+\sin^{2}(\varepsilon_{JC}T)(\ket{e_{1}}\ket{g_{2}}+
\cos(\varepsilon_{JC}T)\ket{g_{1}}\ket{e_{2}})
\cdot(\bra{e_{1}}\bra{g_{2}}+
\cos(\varepsilon_{JC}T)\bra{g_{1}}\bra{e_{2}})
\end{eqnarray}
where, for simplicity, we have fixed the relation $x_{1}
\varepsilon k=x_{2} \varepsilon k\equiv\varepsilon_{JC}$ and same
interaction times for the two atoms, $t_{3}-t_{2}=t_{1}=T$, have
been considered.
Our main result, concerning the effect of the translational
dynamics on the interatomic correlations, can be seen in a simple
way, with a straightforward comparison between the equations
(\ref{rhorid}) and (\ref{rhoridjc}). After a few periods of Rabi
oscillations, the damping factors involved in the scalar products
(\ref{param})(see Eq.(\ref{scalar})), determine a decoherence of
the density matrix (\ref{rhorid}), i.e. the off-diagonal terms go
to zero. As it is known, a system is separable if it is possible
to set the state in the form
$\rho=\sum_{r=1}^{\infty}p_{r}\,W_{r}^{(1)}\otimes W_{r}^{(2)}$,
where $W_{r}^{(1)}$  and $W_{r}^{(2)}$  are states corresponding
to the single subsystem and $p_{r}$ are probabilities
($\sum_{r=1}^{\infty}p_{r}=1$). The Peres-Horodecki test
\cite{Per, Hor2} says that the separability is assured by the
non-negativity of the eigenvalues of the partial transposed matrix
of (\ref{rhorid}). We get for these eigenvalues
\begin{equation}\label{aut4}
\lambda_{1}= c_{1}^{2}P_{2},\quad\quad\lambda_{2}= c_{2}^{2}P_{2},
\quad\quad\lambda_{\pm}=
\frac{P_{1}}{2}(1\pm\sqrt{1+(2|q|c_{1}c_{2}P_{2}\backslash{P_{1})^2}})\quad\quad\quad
\end{equation}
which become all non-negative for $q\rightarrow{0}$. In other
words, from eqs.(\ref{pc}-\ref{scalar}) and subsequent comments,
it is evident that, in our case, the system becomes separable for
interaction times sufficiently large. \emph{Viceversa} a
straightforward application of the the same test to the system
described by (\ref{rhoridjc})  gives the condition
\begin{equation}\label{peresjc}
  \sin^{2}(2 \varepsilon_{JC} T)\sin^{2}(\varepsilon_{JC} T)=0
\end{equation}
for the separability. Evidently, the periodical nature of the
correlations in the JC model context implies an essential non
separability of the system. \\
\begin{figure}[t]
\includegraphics[width = 0.5 \textwidth] {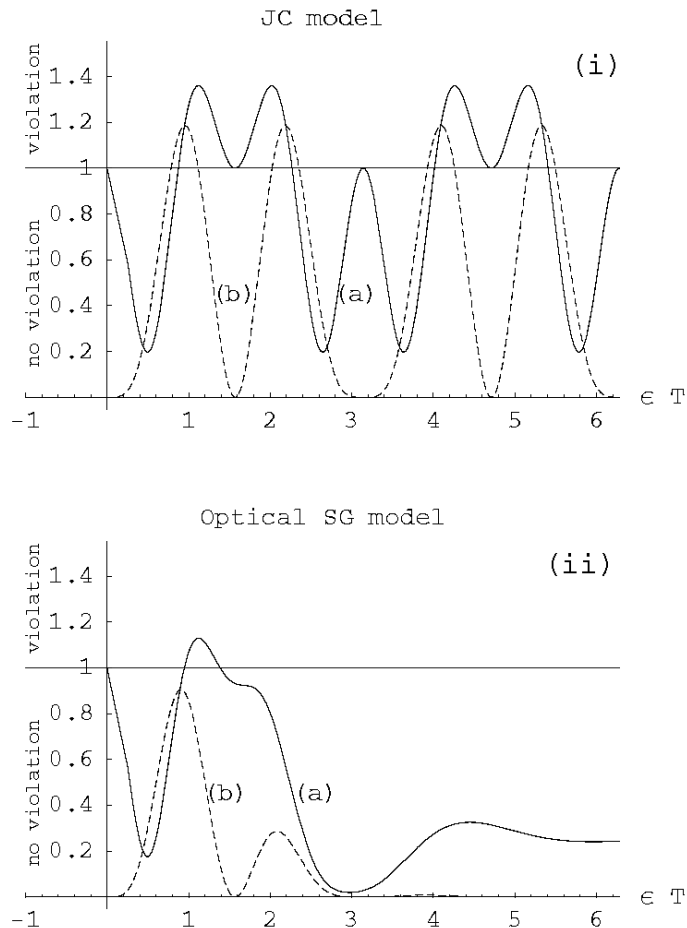} \caption{Graphical solution of Bell's
inequality in terms of the $M(\rho)=\,max\{(a),(b)\}$ function,
for two two-level atoms interacting in succession with the field
of the same cavity. Figure (i) shows the periodicity of
$\nu_{1}+\nu_{2}$ (continuous line)and $2 \nu_{2}$ (dashed line)
when the JC model is adopted and it reproduce the results of
Ref.\cite{Phoe}. Fig. (ii) illustrates the non dissipative damping
of the corresponding quantities due to the entanglement of the
field and the internal atomic variables with the translational
atomic degrees of freedom. Concerning the translational dynamics
we suppose for both the atoms an initial wave packet of minimum
uncertainty, with zero mean value of $\hat{p}_{1}$ and
$\hat{p}_{2}$, centered in $x_{1}=x_{2}=\lambda / 10$ and with a
width $\sigma_{x_{1}}=\sigma_{x_{2}}= \lambda / 10$, where
$\lambda = 2 \pi /k$ is the wavelength of the resonant
\textit{k}-mode of the undamped cavity. The values of the other
parameters are $m=10^{-26}$ kg and $\lambda=10^{-5}$ meters. As
the interaction time increases, the scalar products appearing in
Eq.(\ref{param}) become negligible and the nonlocality of the
quantum correlations between the internal variables of the two
atoms is seriously compromised.} \label{fig1}
\end{figure}
It is also possible to investigate the nature of the interatomic
correlations in terms of the Bell's inequality. Because of its
simplicity, we consider the Horodecki family formulation
\cite{Hor1}, which is equivalent to the standard Clauser, Horne,
Shimony, Holt, (CHSH) formulation \cite{Cla}, when a bipartite
system of spin $1/2$ is involved, as in our case. The test reads:
\emph{A density matrix $\rho$ describing a system composed by two
spin $1/2$  subsystems violates some Bell's inequality in the CHSH
formulation if and only if the relation $M(\rho)>1$ is satisfied}.
The quantity $M(\rho)$  can be defined as follows. Consider the
$3\times3$ matrix $T_{\rho}$ with coefficients
$t_{n,m}=tr(\rho\,\sigma_{n}\otimes \sigma_{m})$, where
$\sigma_{n}$ are the standard Pauli matrices. Diagonalizing the
symmetric matrix $U_{\rho}=T_{\rho}^{T}\cdot T_{\rho}$
($T_{\rho}^{T}$ is the transpose of $T_{\rho}$), and denoting the
three eigenvalues of $U_{\rho}$ by $\nu_{1}$, $\nu_{2}$ and
$\nu_{3}$, then $M(\rho)=
max\{\nu_{1}+\nu_{2},\nu_{1}+\nu_{3},\nu_{2}+\nu_{3}\}$. For the
initial state (\ref{psi0}) there is a degenerate eigenvalue,
$\nu_{2}=\nu_{3}$ and as a consequence $M(\rho)=
max\{\nu_{1}+\nu_{2},2 \nu_{2}\}$. Fig. \ref{fig1} compares the
behaviors of $\nu_{1}+\nu_{2}$ (continuous line) and $2 \nu_{2}$
(dashed line) as a function of the interaction time for the two
models. For simplicity, in both the figures (i) and (ii) we have
assumed $t_{3}-t_{2}=t_{1}=T$ and $t_{2} - t_{1} = T$. The
response of the Bell's inequality test outlines the difference
between the interatomic correlations predicted by the two models.
When the translational dynamics is included in the quantum
treatment of the system, a non dissipative damping of the quantum
correlations appears, which can be ascribed to a which-way
decoherence effect. These effects have actually been extensively
analyzed in other context \cite{Zur,Scull, Bru, Durr, Berte, Tumm,
Chia}, both in theoretical and experimental works. It has here
been shown how the complementarity aspect related to each atom may
weaken the entanglement of an EPR state, leading towards classical
correlations between the two atoms. \\
It is possible, furthermore, to extend the discussion to another
simple case defined by the initial state $
\ket{\psi(0)}=\ket{e_{1}}\ket{g_{2}}\ket{0}\ket{\varphi_{1}(0)}
\ket{\varphi_{2}(0)}$. For this initial state, the quantity
$M(\rho)$ reduces simply to the dashed line of both the figures
(i) and (ii). As it is clear, the violation predicted by the JC
model disappears for all the interaction times, when the
translational dynamics is taken into account. \\
The behaviors shown in Fig. \ref{fig1} suggest that there can be
some difficulties in attaining a violation of Bell's inequality
for massive particles, at least in the conditions here considered.
It is to notice that at the heart of the damping of correlations
there are the scalar products
$\braket{\phi^{+}_{i}}{\phi^{-}_{i}}$ and
$\braket{\phi^{\pm}_{i}}{\varphi_{i}(0)}$, as it is clear from
eq.s (\ref{rhorid}), (\ref{param}) and (\ref{scalar}). As it has
been shown in Ref.\cite{Vag}, similar behaviors of the scalar
products are found when the atoms enter the cavity near an
antinodal point of the field mode. As a consequence it would be
worth to test if our conceptual results can be
extended to the antinodal region. \\
%



\begin{thebibliography}{99}

\bibitem{tel} C.H. Bennett, G. Brassard, C. Cr\'epeau, R. Jozsa,
A. Peres, and W.K. Wootters, {\sl Phys. Rev. Lett.}
{\bf 70}, 1895 (1993); N. Gisin, {\sl Phys. Lett. A}
{\bf 210}, 157 (1996); S.Albeverio, S.-M. Fei, and W.-L. Yang, {\sl Phys. Rev. A} {\bf 66}, 012301 (2002).

\bibitem{Har} J.M. Raimond, M. Brune, and S. Haroche, {\sl Rev. Mod. Phys.}
{\bf 73}, 565 (2001); E. Hagley, X. Ma\^itre, G. Nogues, C.
Wunderlich, M. Brune, J.M. Raimond, and S. Haroche, {\sl Phys.
Rev. Lett.} {\bf 79}, 1 (1997).

\bibitem{Ita} M.A. Rowe, D. Kielpinski, V. Meyer, C.A. Sackett,
W.M. Itano, C. Monroe, and D.J. Wineland,
 {\sl Nature} {\bf 409}, 791 (2001).

\bibitem{Phoe} S.J.D. Phoenix, and S.M. Barnett, {\sl J. Mod. Opt.} {\bf 40}, 979 (1993).

\bibitem{Kni} I.K. Kudryavtsev, and P.L. Knight, {\sl J. Mod.
Opt.} {\bf 40}, 1673 (1993).

\bibitem{Fre} M. Freyberger, P.K. Aravind, M.A. Horne, and A. Shimony, {\sl Phys. Rev. A} {\bf 53}, 1232 (1996).

\bibitem{Zub} S. Qamar, S.-Y. Zhu, and M.S. Zubairy, {\sl Phys. Rev.
A} {\bf 67}, 042318 (2003).

\bibitem{Ray} M.G. Raymer, A.C. Funk, B.C. Sanders, and H. de Guise, {\sl Phys. Rev. A} {\bf 67}, 052104 (2003)

\bibitem{sgo} T. Sleator, T. Pfau, V. Balykin, O. Carnal, and J. Mlynek, {\sl Phys. Rev. Lett.}
{\bf 68}, 1996 (1992); C. Tanguy, S. Reynaud, and C.
Cohen-Tannoudji, {\sl J. Phys. B} {\bf 17}, 4623 (1984); M.
Freyberger, and A. M. Herkommer, {\sl Phys. Rev. Lett.}
{\bf 72}, 1952 (1994); A. Vaglica, {\sl Phys. Rev. A}
{\bf 54}, 3195 (1996).

\bibitem{Asp} A. Aspect, {\sl Nature} {\bf 398}, 189 (1999).

\bibitem{Per} A. Peres, {\sl Phys. Rev. Lett.} {\bf 77}, 1413 (1996).

\bibitem{Wer} R.F. Werner, {\sl Phys. Rev. A} {\bf 40}, 4277 (1989).

\bibitem{Cus} I. Cusumano, A. Vaglica, and G. Vetri, {\sl Phys. Rev. A} {\bf 66}, 043408 (2002).

\bibitem{Aha} Y. Aharonov, D.Z. Albert, and C.K. Au, {\sl Phys. Rev. Lett.}
{\bf 47}, 1029 (1981); R.F. O'Connell, and A.K. Rajagopal, {\sl Phys. Rev. Lett.} {\bf 48}, 525 (1982).

\bibitem{Vag} A. Vaglica, {\sl Phys. Rev. A} {\bf 58}, 3856 (1998).

\bibitem{Chia} M. Chianello, M. Tumminello, A. Vaglica, and G. Vetri, {\sl Phys. Rev. A} {\bf 69}, 053403 (2004).

\bibitem{Hor2} R. Horodecki, P.Horodecki, and M. Horodecki, {\sl Phys.Lett. A} {\bf 223}, 1 (1996).

\bibitem{Hor1} R. Horodecki, P.Horodecki, and M. Horodecki, {\sl Phys.Lett. A} {\bf 200}, 340 (1996).

\bibitem{Cla} J.F. Clauser, M.A. Horne, A. Shimony, and R.A. Holt, {\sl Phys. Rev. Lett.} {\bf 23}, 880 (1969).

\bibitem{Zur} W.H. Zurek, {\sl Physics Today} {\bf 44}, 36 (1991).

\bibitem{Scull} M.O. Scully, B-G. Englert, and H. Walter, {\sl Nature} {\bf 251}, 111 (1991).

\bibitem{Bru} M. Brune, E. Hangley, J. Dreyer, X. Ma\^{i}tre, A. Maali, C. Wunderlich, J.M. Raimond, and S. Haroche, {\sl Phys. Rev. Lett.} {\bf 77}, 4887 (1996).

\bibitem{Durr} S. D\"{u}rr, T. Nonn, and G. Rempe, {\sl Nature} {\bf 395}, 33 (1998).

\bibitem{Berte} P. Bertet, S. Osnaghi, A.Rauschenbeutel, G. Nogues, A. Auffeves, M. Brune, J.M. Raimond, and S. Haroche, {\sl Nature} {\bf 411}, 166 (2001).

\bibitem{Tumm} M. Tumminello, A. Vaglica, and G. Vetri, {\sl Europhys. Lett.} {\bf 65}, 785 (2004).

\end{thebibliography}
\end{document}